\documentclass[12pt]{iopart}
\usepackage{mathptmx}
\usepackage{etoolbox}
\usepackage{amsmath, amsthm, amsfonts}
\newcommand{\D}{\text{D}}
\newcommand{\de}{\text{d}}

\usepackage{cite}

\begin{document}

\title{On the unexpected geometrical origin of the algebra of symmetries}

\author{O. Ram\'irez and Y. Bonder}

\address{Instituto de Ciencias Nucleares\\
Universidad Nacional Aut\'onoma de M\'exico\\
Apartado Postal 70-543, C.P. 04510, Cd. Mx., M\'exico}

\ead{oscar.ramirez@correo.nucleares.unam.mx; bonder@nucleares.unam.mx}
\vspace{10pt}

\begin{abstract}
The fundamental symmetries in gravity and gauge theories, formulated using differential forms, are gauge transformations and diffeomorphisms. These symmetries act in distinct ways on different dynamical fields. Yet, the commutator of these symmetries forms a closed, field-independent algebra. This work uncovers a natural correspondence between this algebra and the Lie bracket of some vector fields on the principal fiber bundle associated with the physical theory, providing a geometric interpretation of the symmetry algebra. Furthermore, we demonstrate that the symmetry algebra is independent of the connection. Finally, we analyze an example illustrating how a specific connection, associated with Lorentz-Lie transformations, simplifies the symmetry algebra in the presence of spacetime Killing vector fields.
\end{abstract}

%% Apr 2021: AIP requests that the corresponding 
%% email to be moved after the affiliations
\makeatletter
\def\@email#1#2{%
 \endgroup
 \patchcmd{\titleblock@produce}
 {\frontmatter@RRAPformat}
 {\frontmatter@RRAPformat{\produce@RRAP{*#1\href{mailto:#2}{#2}}}\frontmatter@RRAPformat}
 {}{}
}%
\makeatother

\section{Introduction}

Fields play a central role in theoretical physics, serving as the fundamental mathematical objects in both classical \cite{Jackson} and quantum theories \cite{Weinberg}. While quantum field theory has achieved remarkable success \cite{PhysRevLett.130.071801}, this paper emphasizes the classical aspects of field theories.

To define any field theory, we need to provide its field content and the associated action functional. The action governs the dynamics of the fields and it also encodes the symmetries of the theory \cite{fradkin_classical_2001,gaset_symmetries_2021}. These symmetries are intimately related with conserved quantities, as shown by Noether's theorems \cite{Noether01011971}. Gauge theories \cite{frampton}, renowned for their success in describing the electromagnetism, weak, and strong interactions, are prime examples of field theories exhibiting a symmetry. In this case the symmetry is associated with Lie groups.

On the other hand, general relativity in its conventional formalism \cite{WaldB}, describes spacetime as a pseudo-Riemannian manifold where the fundamental field is the metric tensor. In this case the symmetry of general relativity is its invariance under diffeomorphisms. In 1922, Cartan extended the metric formalism by introducing an independent connection \cite{cartan1}, triggering the study of numerous alternative gravity theories \cite{Alternative1,Alternative2,Alternative3}. Furthermore, gravity theories can be reformulated using differential forms associated with an internal Lie group, such as the Lorentz group. This approach is referred to as the first-order formalism \cite{Villasenor} and it is particularly useful when the connection is an independent dynamical field.

This paper examines the interplay of symmetries in gauge and gravity theories within the first-order formalism. These theories involve two distinct types of fields: regular fields (algebra-valued $p$-forms) and local connections ($1$-forms with specific gauge transformation properties). Notably, while these fields transform differently under gauge symmetries, they share an intriguing feature: the symmetry algebra is the same regardless of the field type we act upon. This observation suggests that there exists an underlying geometrical structure. Uncovering such a structure is the main goal of this paper. 

The paper is organized as follows. In section~\ref{first order formalism}, we introduce some basic notions within the first-order formalism. Section~\ref{symmetry algebra} is the core of the paper and it presents two derivations of the symmetry algebra: the conventional derivation and a new, geometric method that uncovers a connection to Lie brackets of vector fields on principal fiber bundles. Then, in section \ref{othertransfs}, we show that the structure we find is independent of the connection and we present an example. The conclusions are summarized in Section~\ref{conc}. For completeness, we include basic definitions in \ref{defs}, technical identities in \ref{identities}, and key results on fiber bundles in \ref{PFP}.

Finally, we describe the notation used throughout the paper. Differential forms are employed extensively, omitting spacetime indices. We use the wedge product, exterior derivative, interior derivative, and covariant exterior derivative, we introduce these operations in \ref{defs}. Greek indices $\mu, \nu, \rho, \ldots$ correspond to indexes in the Lie algebra $\mathfrak{g}$, which is associated with the Lie group $G$. Summation is implied for any pair of repeated indices. Moreover, we assume that all the objects we consider are smooth.

\section{Preliminaries}\label{first order formalism}

\subsection{Action}

Our starting point is the construction of an action functional for generic theories. We work in $N$ spacetime dimensions. In addition, we consider a collection of $p$-forms with a algebra index $\Psi^{\mu}$, which represent general matter fields and/or vielbeins. We employ vielbeins to describe gravity, which are more commonly known as tetrads in the specific case of four-dimensional spacetime \cite[chapter 3.4b]{WaldB}. We also work with an arbitrary collection of local connection/gauge fields, with an arbitrary internal Lie group; we denote the local connection fields (or simply gauge fields) by $\mathcal{A^{\mu}_{\ \nu}}$. Note that, to simplify the notation, we omit the index specifying which field/connection within the collection is used, as done in \cite{BonCris}. Note that this formalism is extremely general in that it can describe a wide range of theories.

To study the dynamics and symmetries of a theory, we need to provide an action. By definition, the action is a spacetime integral of an $N$-form called the Lagrangian, $L$. In turn, $L$, which depends on all relevant fields, must transform as a scalar under gauge transformations. To implement this condition, we assume that $L$ has no free algebra indexes and that it only depends on the local connections through its curvature, $F^{\mu}_{\ \nu}$, and the covariant exterior derivative of $\Psi^{\mu}$, $\D\Psi^{\mu}$; these objects are defined in \ref{defs}. With these considerations, a generic action takes the form
\begin{equation}
S[\Psi^{\mu},\mathcal{A}^{\mu}_{\ \nu}] = \int \ L[\Psi^{\mu}, \D\Psi^{\mu} ,F^{\mu}_{\ \nu} ].
\label{Acc}
\end{equation}

Observe that we do not explicitly consider higher-derivative terms in our analysis. Naively, one might argue that such terms are unnecessary since $\D^2\Psi^{\mu} = F^{\mu}_{\ \nu}\Psi^{\nu}$ and $\D F^{\mu}_{\ \nu} = 0$ (see equation \eqref{Bianchi}). To incorporate higher-order derivatives into the equations of motion, we can employ a Hodge star map \cite{Naka}, as argued in the context of the Lovelock theorem \cite{Hassaine}. The Hodge star map, in turn, necessitates the introduction of a (background or dynamical) spacetime metric. In this work, we do not explicitly include such higher-order theories. Nonetheless, their incorporation into our framework is straightforward.

We can express the variation of $S$, as given in \eqref{Acc}, as
\begin{eqnarray}
\delta S &=& \int \left[\delta \Psi^{\mu} \wedge \frac{\partial L}{\partial \Psi^{\mu}} + \delta \D \Psi^{\mu} \wedge \frac{\partial L}{\partial \D \Psi^{\mu}} + \delta F^{\mu}_{\ \nu} \wedge \frac{\partial L}{\partial F^{\mu}_{\ \nu} }\right] \\
 &=& \int\left\{ \delta \Psi^{\mu} \wedge \left[ \frac{\partial L}{\partial \Psi^{\mu}} - (-1)^{p} \D \left( \frac{\partial L}{\partial \D \Psi^{\mu}}\right)\right] \right.\nonumber\\
&& \left.+ \delta \mathcal{A}^ {\mu}_{\ \nu} \wedge \left[ \Psi^{\nu} \wedge \frac{\partial L}{\partial \D \Psi^{\mu}} + \D \left(\frac{\partial L}{ \partial F^{\mu}_{\ \nu}}\right)\right]\right\}\nonumber \\ 
&&+\int \de \left(\delta \Psi^{\mu} \wedge \frac{\partial L}{\partial \D \Psi^{\mu}} + \delta \mathcal{A}^{\mu}_{\ \nu} \wedge \frac{\partial L}{ \partial F^{\mu}_{\ \nu}}\right),
\label{variS}
\end{eqnarray}
where we use $ \delta F^{\mu}_{\ \nu} = \D \delta \mathcal{A}^{\mu}_{\ \nu}$ and $\delta \D \Psi^ {\mu} = \D \delta \Psi^{\mu} + \delta \mathcal{A}^{\mu}_{\ \nu} \wedge \Psi^{\nu}$, and the (graded) Leibniz rule for $\D$. We can read off the equations of motion directly from equation \eqref{variS}. However, these equations are not required for the analysis that follows. Instead, we shift our focus to the symmetries of the theory.

\subsection{Symmetries}

An infinitesimal transformation of the fields, $\delta \Psi^{\mu}$ and $\delta \mathcal{A}^ {\mu}_{\ \nu}$, is a symmetry of the theory if, when inserted into \eqref{variS} produces ${\delta}S=0$\footnote{A transformation is a pseudosymmetry if there exists a form $B$ of co-dimension $1$ such that ${\delta}S = \int \de B$; since the distinction between symmetries and pseudosymmetries is irrelevant for this study, we ignore this distinction.}. For instance, in a gauge theory, gauge transformations are considered symmetries of the theory. Moreover, in the absence of background structures \cite{BonCris}, diffeomorphisms are also symmetries of the theory.

An infinitesimal gauge transformation associated with $\lambda \in \mathfrak{g}$, when acting on a regular field, takes the form
\begin{equation}\label{GT}
\delta_{\text{GT}}(\lambda) \Psi^{\mu} = - \lambda^{\mu}_{\ \nu} \Psi^{\nu}.
\end{equation}
In addition, for $\text{D} \Psi^{\mu}$ to transform covariantly, the infinitesimal transformation for the local connection is
\begin{equation}
\delta_{\text{GT}}(\lambda) \mathcal{A}^{\mu}_{\ \nu} = \text{D} \lambda^{\mu}_{\ \nu}.
\label{GTc}
\end{equation}
On the other hand, infinitesimal diffeomorphisms act on the fields by the Lie derivative along the vector field $\xi$ which generates this transformation. We denote this derivative as $\mathcal{L}_{\xi}$. Thus, the fields' transformations associated with a diffeomorphism are
\begin{eqnarray}
\delta_{\text{Diff}}(\xi) \Psi^{\mu} &=& \mathcal{L}_{\xi} \Psi^{\mu} = \text{i}_{\xi} \text{d} \Psi^{\mu} + \text{d} \text{i}_{\xi} \Psi^{\mu},
\label{diffpsi}
\\
\delta_{\text{Diff}}(\xi) \mathcal{A}^{\mu}_{\ \nu} &=& \mathcal{L}_{\xi} \mathcal{A}^{\mu} _{\ \nu} = \text{i}_{\xi} \text{d} \mathcal{A}^{\mu}_{\ \nu} + \text{d} \text{i}_ {\xi} \mathcal{A}^{\mu}_{\ \nu},
\label{diffcon}
\end{eqnarray}
where we use Cartan's magic formula \eqref{magica}.

Note that, when acting on objects with algebra indexes, diffeomorphisms are not necessarily covariant, namely, $\delta_{\text{Diff}}(\xi) \Psi^{\mu}$ does not transform according to equation \eqref{GT}. In fact, something analogous occurs with the connection. One could have guessed this result from the fact that the corresponding infinitesimal transformation contains exterior derivatives, and not covariant exterior derivatives $\D$, which, as the name suggests, are built to transform covariantly under gauge transformations. Hence, it is convenient to introduce the notion of a local translation (LT) (or covariant diffeomorphisms, as they are known in \cite{Simone}). Infinitesimally, an LT associated with $\xi$ acts as
\begin{eqnarray}
\delta_{\text{LT}}(\xi) \Psi^{\mu} &=& \text{i}_{\xi} \text{D} \Psi^{\mu} + \text{D} \text{i}_{\xi} \Psi^{\mu},
\label{LTpsi}
\\
\delta_{\text{LT}}(\xi) \mathcal{A}^{\mu}_{\ \nu} &=& \text{i}_{\xi} F^{\mu}_{\ \nu}.
\label{LTcon}
\end{eqnarray}
We can readily show that any LT satisfies \cite{BonCris}
\begin{equation}\label{LTasDiffandGauge}
\delta_{\text{LT}}(\xi) = \delta_{\text{Diff} }(\xi) - \delta_{\text{GT}} (\text{i}_{\xi} \mathcal{A}).
\end{equation}
In other words, any LT is a combination of a diffeomorphism and a gauge transformation with a field-dependent gauge parameter $\text{i}_{\xi} \mathcal{A}$. Importantly, equation \eqref{LTasDiffandGauge} follows irrespectively of whether we act on $\Psi^{\mu}$ or $\mathcal {A}^{\mu}_{\ \nu}$. Therefore, any theory that is symmetric under gauge transformation and diffeomorphisms, is also symmetric under LT. Having defined the symmetries, we can proceed to calculate the symmetry algebra.

\section{Symmetry algebra}\label{symmetry algebra}

In this section, we calculate the commutators of the symmetries we describe above and show that they form a closed and field-independent algebra. We define the commutator of any transformations associated with the infinitesimal generators $\eta_{1}$ and $\eta_{2}$ in the natural form\footnote{We use a different definition for the commutator than \cite{BonCris,kiriushcheva2009hamiltonian,Monte}.}:
\begin{equation}
[\delta (\eta_{1}), \delta (\eta_{2}) ]= \delta (\eta_{1}) \delta (\eta_{2})- \delta (\eta_{2})\delta (\eta_{1}).
\label{conmSS}
\end{equation}
Keep in mind that these transformations act differently on regular fields and connections. However, once we apply one of these transformations, the resulting object transforms like a regular field.

The commutators can be calculated using two approaches: the conventional method, which involves acting with the symmetries on fields, or the geometrical method we propose below. We begin this section by presenting the conventional calculation. 

\subsection{Conventional calculation}

In this part of the paper, we compute the commutators of gauge transformations and diffeomorphisms by acting on the fields. We begin by computing the commutators when acting on a connection field. To obtain the commutator of two gauge transformations, we note that successive gauge transformations produce
\begin{eqnarray}
\delta_{\text{GT}} (\lambda_{1})\delta_{\text{GT}} (\lambda_{2}) \mathcal{A}^{\mu}_{\ \nu} &=&\delta_{\text{GT}} ( \lambda_{1})\{ \de (\lambda_{2})^{ \mu}_{\ \nu} +\mathcal{A}^{\mu}_{\ \rho}(\lambda_{2})^{\rho}_{\ \nu}- \mathcal{A}^{\rho}_{\ \nu}(\lambda_{2} )^{\mu}_{\ \rho} \} \nonumber\\ 
&=&
 -(\D [\lambda_{1},\lambda_{2}]_{\mathfrak{g}} )^{\mu}_{\ \nu} + (\D \lambda_{1})^{\mu}_{\ \rho} (\lambda_{2})^{\rho}_{\ \nu} - (\D \lambda_{1})^{\rho}_{\ \nu} (\lambda_{2})^{\mu}_{\ \rho} ,\nonumber\\
 && 
\label{gtalgebra3}
\end{eqnarray}
where, $\D \lambda^{ \mu}_{\ \nu} =\de \lambda^{ \mu}_{\ \nu} +\mathcal{A}^{\mu}_{\ \rho}\lambda^{\rho}_{\ \nu}- \mathcal{A}^{\rho}_{\ \nu}\lambda^{\mu}_{\ \rho}$ (see equation \ref{def D}) and $[\cdot,\cdot]_{\mathfrak{g}}$ is the Lie algebra commutator. Hence,
\begin{equation}
[\delta_{\text{GT}} (\lambda_{1}),\delta_{\text{GT}} (\lambda_{2})] \mathcal{A}^{\mu}_{\ \nu} =- (\D [\lambda_{1},\lambda_{2}]_{\mathfrak{g}} )^{\mu}_{\ \nu} = \delta_{\text{GT}} (-[\lambda_{1},\lambda_{2}]_{\mathfrak{g}})\mathcal{A}^{\mu}_{\ \nu},
\label{gtalgebra4}
\end{equation}
where we use equation \eqref{GT} in the last step. Equation \eqref{gtalgebra4} reproduces the Lie algebra (up to a sign that we discuss below), as expected.

Our next task is to compute the commutator of a gauge transformation and an LT. We first note that
\begin{eqnarray}
\delta_{\text{LT}} (\xi) \delta_{\text{GT}} (\lambda) \mathcal{A}^{\mu}_{\ \nu} 
&=& \D \text{i}_{\xi} \D \lambda^{\mu}_{\ \nu} + \text{i}_{\xi} \D \D \lambda^{\mu}_{\ \nu} \nonumber\\ 
&=& \D( \text{i}_{\xi} \D \lambda^{\mu}_{\ \nu}) + \text{i}_{\xi} F^{\mu}_{\ \rho} \lambda^{\rho}_{\ \nu} - \text{i}_{\xi}F^{\rho}_{\ \nu} \lambda^{\mu}_{\ \rho},
\label{gtalgebra6}
\end{eqnarray}
where we use equation \eqref{Dsquared}. Conversely,
\begin{equation}
\delta_{\text{GT}} (\lambda)\delta_{\text{LT}} (\xi) \mathcal{A}^{\mu}_{\ \nu} = \delta_{\text{GT}} (\lambda) \text{i}_{\xi} F^{\mu}_{\ \nu} = - \text{i}_{\xi}F^{\rho}_{\ \nu} \lambda^{\mu}_{\ \rho} + \text{i}_{\xi} F^{\mu}_{\ \rho} \lambda^{\rho}_{\ \nu}.
\label{gtalgebra5}
\end{equation}
Thus,
\begin{equation}
[\delta_{\text{LT}} (\xi) ,\delta_{\text{GT}} (\lambda) ]\mathcal{A}^{\mu}_{\ \nu} = \D( \text{i}_{\xi} \D \lambda^{\mu}_{\ \nu}) = \delta_{\text{GT}} (\text{i}_{\xi} \D \lambda) \mathcal{A}^{\mu}_{\ \nu}.
\label{gtalgebra7}
\end{equation}
In other words, the commutator of a gauge transformation and an LT is a gauge transformation with parameter $(\text{i}_{\xi} \D \lambda)^{\mu}_{\ \nu} $.

Now we study the commutator of two LTs. We observe that
\begin{eqnarray}
\delta_{\text{LT}} (\xi_{1}) \delta_{\text{LT}} (\xi_{2}) \mathcal{A}^{\mu}_{\ \nu}
&=& \delta_{\text{LT}} (\xi_{1}) \text{i}_{\xi_{2}} F^{\mu}_{\ \nu} \nonumber\\
&=& \D \text{i}_{\xi_{1}} \text{i}_{\xi_{2}} F^{\mu}_{\ \nu} +\text{i}_{\xi_{1}} \D \text{i}_{\xi_{2}} F^{\mu}_{\ \nu}.
\end{eqnarray}
%\begin{equation}
%\begin{split}
%\delta_{\text{LT}} (\xi_{1}) \delta_{\text{LT}} (\xi_{2}) W^{\mu}_{\ \nu} &= \delta_{\text{LT}} (\xi_{1}) \text{i}_{\xi_{2}} F^{\mu}_{\ \nu}= \xi_{2}^{ \ \alpha} ( D \text{i}_{\xi_{1}} \text{i}_{\alpha} F^{\mu}_{\ \nu} + \text{i}_{\xi_{1}}D \text{i}_{\alpha} F^{\mu}_{\ \nu}) \\
%& = \xi_{2}^{\ \alpha} (\mathcal{L}_{\xi_{1} } \text{i}_{\alpha} F^{\mu}_{\ \nu} - \text{i}_{\xi_{1}} W^{\beta}_{\ \alpha} \text{i}_{\beta} F^{\mu}_{\ \nu} +\text{i}_{\xi_{1}} W^{\mu}_{\ \beta} \text{i}_{\alpha} F^{\beta}_{\ \nu} - \text{i}_{\xi_{1}} W^{\beta}_{\ \nu} \text{i}_{\alpha} F^{\mu}_{\ \beta} ).
%\end{split}
%\label{gtaalgebra13}
%\end{equation}
%It is easy to prove the following identity:
%\begin{equation}
%\begin{split}
%\D(\text{i}_{\xi_{1}} \text{i}_{\xi_{2}} F^{\mu}_{\ \nu}) = \mathcal{L}_{\xi_{1}} \text{i}_{\xi_{2}} F^{\mu}_{\ \nu} -\text{i}_{\xi_{1}} \mathcal{L}_{\xi_{2}} F^{\mu}_{\ \nu} - \text{i}_{\xi_{2}} W^{\mu}_{\ \alpha} \text{i}_{\xi_{1}} F^{\alpha}_{\ \nu} &+ \text{i}_{\xi_{1}} W^{\mu}_{\ \alpha} \text{i}_{\xi_{2}}F^{\alpha}_{\ \nu} \\
%&+ \text{i}_{\xi_{2}} W^{\alpha}_{\ \nu} \text{i}_{\xi_{1}} F^{\mu}_{\ \alpha} - \text{i}_{\xi_{1}} W^{\alpha}_{\ \nu} \text{i}_{\xi_{2}} F^{\mu}_{\ \alpha}.
%\end{split}
%\label{gtalgebra14}
%\end{equation}
From equation \eqref{identidad2tensor} and using the Bianchi identity \eqref{Bianchi}, we obtain
\begin{eqnarray}
[\delta_{\text{LT}} (\xi_{1}), \delta_{\text{LT}} (\xi_{2}) ]\mathcal{A}^{\mu}_{\ \nu} 
&=& \D \text{i}_{\xi_{1}} \text{i}_{\xi_{2}} F^{\mu}_{\ \nu} +\text{i}_{\xi_{1}} \D \text{i}_{\xi_{2}} F^{\mu}_{\ \nu} -\text{i}_{\xi_{2}} \D \text{i}_{\xi_{1}} F^{\mu}_{\ \nu} \nonumber\\
&&- \text{i}_{\xi_{2}} \text{i}_{\xi_{1}} \D F^{\mu}_{\ \nu} - \D \text{i}_{\xi_{2}} \text{i}_{\xi_{1}} F^{\mu}_{\ \nu} \nonumber\\
&=& \text{i}_{[\xi_{1},\xi_{2}]_{M}} F^{\mu}_{\ \nu} -\D \text{i}_{\xi_{2}} \text{i}_{\xi_{1}} F^{\mu}_{\ \nu} \nonumber\\
&=& \delta_{\text{LT}}([\xi_{1},\xi_{2}]_{M}) \mathcal{A}^{\mu}_{\ \nu} -\delta_{\text{GT}} (\text{i}_{\xi_{2}} \text{i}_{\xi_{1}} F^{\mu}_{\ \nu}) \mathcal{A}^{\mu}_{\ \nu}.
\label{LTalgebra235}
\end{eqnarray}
Hence, the commutator of two LT is another LT whose parameter is the Lie bracket of the two (spacetime) vector fields generating the LTs, $[\xi_{1}, \xi_{2}]_{M}$, plus a gauge transformation whose parameter is the curvature $2$-form evaluated on $\xi_{1}$ and $\xi_{2}$.

We turn to calculate the same commutators but now acting on $\Psi^{\mu}$. We can verify that
\begin{equation}
\delta_{\text{GT}} (\lambda_{1})\delta_{\text{GT}} (\lambda_{2}) \Psi^{\mu} = -\delta_{\text{GT}} (\lambda_{1}) (\lambda_{2})^{\mu}_{\ \nu} \Psi^{\nu} =(\lambda_{1})^{\mu}_{\ \rho} (\lambda_{2})^{\rho}_{\ \nu} \Psi^{\nu}.
\label{gtalgebra}
\end{equation}
Thus,
\begin{equation}
[\delta_{\text{GT}} (\lambda_{1}),\delta_{\text{GT}} (\lambda_{2})] \Psi^{\mu} 
= ([\lambda_{1}, \lambda_{2}]_{\mathfrak{g}})^{\mu}_{\ \nu} \Psi^{\nu}
= \delta_{\text{GT}} (-[\lambda_{1}, \lambda_{2}]_{\mathfrak{g}})\Psi^{\mu} .
\label{gtalgebra2}
\end{equation}
We now proceed to calculate the commutator of a gauge transformation and an LT when acting on $\Psi^{\mu}$. Notice that
\begin{eqnarray}
\delta_{\text{LT}} (\xi) \delta_{\text{GT}} (\lambda) \Psi^{\mu} 
&=&- \D \text{i}_{\xi} ( \lambda^{\mu}_{\ \nu} \Psi^{\nu}) - \text{i}_{\xi} \D ( \lambda^{\mu}_{\ \nu} \Psi^{\nu})\nonumber \\
&=& -\D \lambda^{\mu}_{\ \nu} \wedge \text{i}_{\xi} \Psi^{\nu} - \lambda^{\mu}_{\ \nu} \D \text{i}_{\xi} \Psi^{\nu} - \text{i}_{\xi} (\D \lambda^{\mu}_{\ \nu} \wedge \Psi^{\nu} + \lambda^{\mu}_{\ \nu} \D \Psi^{\nu} )\nonumber\\
&=&- \lambda^{\mu}_{\ \nu} \D \text{i}_{\xi} \Psi^{\nu} - \lambda^{\mu}_{\ \nu} \text{i}_{\xi} \D \Psi^{\nu} - \text{i}_{\xi} \D \lambda^{\mu}_{\ \nu} \Psi^{\nu}.
\label{gtaalgebra11}
\end{eqnarray}
When acting on the opposite order, we get
\begin{equation}
\delta_{\text{GT}} (\lambda) \delta_{\text{LT}} (\xi) \Psi^{\mu} = \delta_{\text{GT}} (\lambda) ( \D \text{i}_{\xi} \Psi^{\mu} + \text{i}_{\xi} \D \Psi^{\mu}) = -\lambda^{\mu}_{\ \nu} \D \text{i}_{\xi} \Psi^{\nu } - \lambda^{\mu}_{\ \nu} \text{i}_{\xi} \D \Psi^{\nu}.
\label{gtalgebra9}
\end{equation}
Therefore, the commutator is given by
\begin{equation}
[\delta_{\text{LT}} (\xi) ,\delta_{\text{GT}} (\lambda) ]\Psi^{\mu} =- \text{i}_{\xi} \D \lambda^{\mu}_{\ \nu} \Psi^{\nu} = \delta_{\text{GT}} (\text{i}_{\xi} \D \lambda) \Psi^{\mu}.
\label{gtalgebra12}
\end{equation}
Finally, we obtain the commutation of two LT when acting on $\Psi^{\mu}$. Again, we begin by calculating 
\begin{equation}
\delta_{\text{LT}} (\xi_{1}) \delta_{\text{LT}} (\xi_{2}) \Psi^{\mu} 
= \D \text{i}_{\xi_{1}} \D \text{i}_{\xi_{2}} \Psi^{\mu} + \text{i}_{\xi_{1}} \D \D \text{i}_{\xi_{2}} \Psi^{\mu} + \D \text{i}_{\xi_{1}} \text{i}_{\xi_{2}} \D\Psi^{\mu} + \text{i}_{\xi_{1}} \D \text{i}_{\xi_{2}} \D\Psi^{\mu}.
\label{gtalgebra16}
\end{equation}
By utilizing equations \eqref{iden5} and \eqref{iden6}, we can calculate the commutator, which is expressed as
\begin{eqnarray}
[\delta_{\text{LT}} (\xi_{1}), \delta_{\text{LT}} (\xi_{2}) ]\Psi^{\mu} 
&=& \D \text{i}_{[\xi_{1}, \xi_{2}]_{M}} \Psi^{\mu} + \text{i}_{[\xi_{1}, \xi_{2}]_{M}} \D \Psi^{\mu} + \text{i}_{\xi_{2}}\text{i}_{\xi_{1}} F^{\mu}_{\ \nu} \Psi^{\nu}\nonumber \\
&=& \delta_{\text{LT}} ([\xi_{1} ,\xi_{2}]_{M}) \Psi^{\mu} - \delta_{\text{GT}} (F(\xi_{1} , \xi_{2})) \Psi^{\mu},
\label{iden7}
\end{eqnarray}
where we interchangeably use $\text{i}_Y \text{i}_X F $ and $ F(X,Y)$.

Upon inspection of the commutators, we observe that, despite the strong dependence of the calculations on the specific field types, they form a closed algebra that is field independent. This algebra has the following structure:
\begin{subequations}\label{algebra}
\begin{eqnarray}
\left[\delta _{\text{GT}} (\lambda _{1}),\delta _{\text{GT}} (\lambda _{1})\right] &=& - \delta _{\text{GT}} ([\lambda_{1},\lambda_{2}]_{\mathfrak{g}})
\label{gtgt}
\\
\left[\delta _{\text{LT}} (\xi),\delta _{\text{GT}} (\lambda)\right] &=& \delta _{\text{GT}} (\text{i}_{\xi}\D \lambda )
\label{ltgt}
\\
\left[\delta_{\text{LT}} (\xi_{1}),\delta_{\text{LT}} (\xi_{2})\right] &=& \delta_{\text{LT}} ( [\xi_{1}, \xi_{2}]_{\mathfrak{g}} ) - \delta_{\text{GT}} (F(\xi_{1} , \xi_{2})).
\label{ltlt}
\end{eqnarray}
\end{subequations}
The surprising field-independence of this algebra strongly suggests the existence of an underlying mathematical structure. Uncovering such a structure is the focus of the next subsection.

\subsection{Geometric method}\label{geometric}

In this subsection, we explore the connection between the symmetry space of a given physical theory and a principal fiber bundle. This approach is motivated by fact that in a principal fiber bundle all fields have identical transformation properties. In contrast, the diverse transformations observed in spacetime arise as a consequence of the projection onto the base manifold (see \ref{PFP}).

Our method is begins by establishing a homomorphism between two distinct vector spaces. We now proceed to construct a vector space associated with the symmetries of the theories.

\subsubsection{Symmetry vector space}

The first important observation is that a gauge transformation and an LT\footnote{It is relevant to point that, according to the terminology of \cite{Lee}, these are field-independent transformations.} define a real vector space $\mathcal{V}$. Let $\eta$ and $\chi$ be infinitesimal transformation parameters. These parameters belong to a vector space that is the external direct sum of two vector spaces: the set of spacetime vector fields and the Lie algebra $\mathfrak{g}$.

We define the addition of the transformations $\delta (\eta), \delta (\chi)\in \mathcal{V}$ as
\begin{equation}
 [\delta (\eta) + \delta (\chi)] \Theta = \delta (\eta+\chi) \Theta.
 \label{defsum}
\end{equation}
where $\Theta$ represents any dynamical field, namely, a regular field or a local connection, possibly with algebra indexes. This addition is commutative and associative. Also, by construction, this addition is independent of the field we act on. Analogously, we define the scalar multiplication as
\begin{equation}
 [\alpha\delta(\eta)] \Theta = \delta(\alpha\eta) \Theta,
 \label{scalarprodd}
\end{equation}
which is also well-defined and field-independent.

The identity element in $\mathcal{V}$ is the transformation with parameter $\eta=0$, which corresponds to the zero vector field in spacetime and the zero element in $\mathfrak{g}$. The inverse of $\delta(\eta)$ is $\delta(-\eta)$ where again we use the inverse of each vector space. We can verify the compatibility of scalar multiplication, the existence of the identity element of scalar multiplication, and the distributivity of scalar multiplication for vector addition and field addition. Thus, equipped with these operations, $\mathcal{V}$ is a real vector field.

\subsubsection{Symmetries as vector fields on a principal fiber bundle}\label{mapeo}

A principal fiber bundle is a manifold, $P$, that is locally $M\times G$, where $M$ is the base manifold and $G$ is a Lie group \cite{Prabhu_2017,2018arXiv180205345B,Fatibene2011,ALEKSEEVSKY1995371}. $P$ is also equipped with a connection. This connection separates directions (i.e., tangent vectors) into ``vertical'', that is, along a fiber, which is diffeomorphic to $G$ \cite{Naka}, and ``horizontal.'' More concretely, given a connection, we can split the tangent space over $u \in P$, $T_{u}P$, into its vertical and horizontal parts, denoted respectively by $V_{u}P$ and $H_{u}P$, in such a way that $T_{u}P=V_{u}P \bigoplus H_{u}P$. For more details about the construction of a principal fiber bundle, we refer our readers to \ref{PFP}.

The goal of this part is to establish a relation between the field transformations and $P$. In our context, the base manifold $M$ corresponds to spacetime and $G$ is the Lie group associated with the theory's gauge symmetry. Hence, vertical directions on $P$ are related to gauge transformations.
%while horizontal directions involve changes in the spacetime points. We can thus encode in the connection (whose projection in $M$ corresponds to the local connection $\mathcal{A}$) the specific transformation under consideration, namely, the particular combination of gauge transformation and diffeomorphisms that we use. For instance, a connection can be associated with an LT, which in turn is a combination of gauge transformations and diffeomorphisms, as we show in \eqref{LTasDiffandGauge}. This is the connection we use in the during the rest of this section. Evidently, different combinations of transformations call for different connections; this is discussed in section \ref{othertransfs}.

Recall that group multiplication on the right by $g\in G$ can be thought of as a map $R_g:G \to G$, and thus, its pushforward, ${R_g}_*$, is a map between tangent spaces; of course, something analogous could be done with the left multiplication. As such, ${R_g}_*$ allows us to compare vectors on different points of $P$ within a single fiber. Moreover, we can construct a vector field on a fiber from a single vector by requiring invariance under the action of ${R_g}_*$ for all $g \in G$.

What is more, an element of $\mathfrak{g}$ can be interpreted as a vector on $P$ given that $\mathfrak{g} \cong T_e G$, where $e \in G$ is the identity element \cite{Naka}. Consider a map that assigns, to each spacetime point $x \in M$, an element of $\mathfrak{g}$, denoted $\lambda(x)$. By treating $\lambda(x)$ as a vector at the identity in the fiber associated with $x$, for all $x \in M$, and by extending this vector on the corresponding fibers through invariance under right multiplication, we obtain a vector field throughout $P$, denoted by $X_\lambda$. We refer to $X_\lambda$ as the right-invariant vector field corresponding to $\lambda(x)$.

Using the definition of a vertical vector in terms of the pushforward of the projector map $\pi:P\to M$, it is straightforward to show that $X_\lambda$ is a vertical vector field on $P$. Moreover, the zero element of $\mathfrak{g}$ produces the trivial vector field on $P$; we can use this fact to show that the map $\lambda(x)\to X_\lambda$ is onto.

In the particular case when $\lambda(x) = \lambda$ is constant throughout spacetime, we can utilize its associated fundamental vector field, $\lambda^{\#}$, which is a vertical and left-invariant vector field over the fiber (see equation \ref{PFP}). Notably, in this case, it can be shown that \cite{Werner}
\begin{equation}
 X_{\lambda} = {R_g}_* \lambda^{\#} = [\text{Ad}_{g^{-1}}\lambda]^{\#},
 \label{fundamentalRI}
\end{equation}
where $\text{Ad}_{g^{-1}}$ corresponds to the adjoint map, which we define in \ref{PFP}.

At this stage, it is clear that we should represent a gauge transformation, $\delta_{\text{GT}}(\lambda)$, by the vertical vector field $X_\lambda$. But how can we represent an LT in $P$? Using similar arguments to those described in \ref{PFP}, we can associate to any vector field on $M$, $\xi$, a unique \cite{Werner} horizontal vector field on $P$, denoted by $\tilde{\xi}$, which is called the horizontal lift of $\xi$. Naturally, this horizontal lift should correspond to $\delta_{\text{LT}}(\xi)$. We turn to show that there is a homomorphism linking $\mathcal{V}$ with these vector fields on $P$.

\subsubsection{Homomorphism}

We proceed to use the associations of symmetries and vector fields on $P$ we describe above to construct a homomorphism $\mathcal{V} \rightarrow \mathfrak{X}(P)$ where $\mathfrak{X}(P)$ is the set of vector fields on $P$. Let $\widetilde{\Phi}:\mathcal{V} \rightarrow \mathfrak{X}(P)$ be such that it takes $\delta_{\text{GT}}(\lambda)$ to $X_\lambda$ and $\delta_{\text{LT}}(\xi)$ to $\tilde{\xi}$. 
%In that sense, $\Phi$ correspond to a linear transformation between this two vector spaces. We are going to discuss some properties of the map $\Phi$ below but, first we will look the relevant vector fields that we use on this work.
In addition, we require $\widetilde{\Phi}$ to preserve the addition and to be linear in the following sense
\begin{equation}
 \widetilde{\Phi}(\alpha\delta(\eta)+ \beta\delta(\chi)) = \alpha\widetilde{\Phi}(\delta(\eta) ) + \beta\widetilde{\Phi}(\delta(\chi)).
 \label{PhiSV}
\end{equation}
By construction, $\widetilde{\Phi}$ is injective, but it is not surjective, as there are vector fields on $P$ that do not correspond to the symmetries under consideration. Since we require an invertible map, we work with $\Phi$, which is the restriction of $\widetilde{\Phi}$ such that its codomain is $\mathfrak{X}_{\rm r}(P) \subset \mathfrak{X}(P)$, where $\mathfrak{X}_{\rm r}(P) $ is the union of the set of right-invariant vector fields and the set of horizontal lifts. With this restriction, it is clear that $\Phi$ is a homomorphism. In the following sections, we make extensive use of $\Phi$. However, for convenience and since it does not cause any ambiguity, from this point onward we omit writing $\Phi$ explicitly.

We have a homomorphism that allows us to associate gauge transformations and LT with vector fields on $P$. The natural candidate to generate the symmetry algebra is the Lie bracket of such vector fields. We now demonstrate that this is indeed the case.

\subsubsection{Lie brackets}

We begin this part of the paper by calculating the Lie bracket of two horizontal lifts. Let $X_{\lambda_1}$ and $X_{\lambda_2}$ be two such fields. The calculation can be carried out by inserting these fields into the curvature on $P$, denoted by $\Omega$. By definition (see \ref{PFP}), this curvature vanishes when acting on vertical vector fields, leading to $\Omega(X_{\lambda_1}, X_{\lambda_2}) = 0$. Moreover, using equation \eqref{Cse}, we obtain
\begin{eqnarray}
\omega[X_{\lambda_{1}},X_{\lambda_{2}}]_{P} &=& \Omega(X_{\lambda _{1}},X_{\lambda _{2}})+\omega[X_{\lambda_{1 }},X_{\lambda_{2}}]_{P}\nonumber \\
&=& \de_{P} \omega (X_{\lambda _{1}},X_{\lambda _{2}}) + [\text{Ad}_{g^{-1}} \lambda _{1}( x),\text{Ad}_{g^{-1}} \lambda_{2}(x)]_{\mathfrak{g}}+\omega[X_{\lambda_{1) }},X_{\lambda_{2}}]_{P} \nonumber\\
 &=& X_{\lambda_{1}}[g^{-1} \lambda _{2}(x) g]- X_{\lambda_{2}}[g^{-1} \lambda _{1}(x) g] +\text{Ad}_{g^{-1}} ( [ \lambda_{1} (x) , \lambda_{2} (x)]_{\mathfrak {g}}) \nonumber\\
 &=& - \text{ Ad}_{g^{-1}} ( [ \lambda_{1} (x) , \lambda_{2} (x)]_{\mathfrak{g}}) ,
\label{curvgtgt2}
\end{eqnarray}
where $[\cdot,\cdot]_{P}$ is the Lie bracket in $P$, and we use
\begin{equation}
\omega(X_{\lambda}) = \text{Ad}_{g^{-1}} [\lambda(x)],
\end{equation}
and
\begin{equation}
X_ {\lambda _{1}}[\text{Ad}_{g^{-1}} [\lambda _{2}(x)]] = - \text{Ad}_{g^{-1}} ( [ \lambda_{1} (x) , \lambda_{2} (x)]_{\mathfrak{g}}).\label{omegahorhor}
\end{equation}
Notice that, for notational simplicity, we suppress Greek indices whenever possible. By applying the pushforward of the projection map, $\pi$, to equation \eqref{omegahorhor}, we can show that $[X_{\lambda _{1}},X_{\lambda _{2}}]_{P}$ does not have a horizontal part. Thus,
\begin{equation}
[X_{\lambda _{1}},X_{\lambda _{2}}]_{P} = - [\text{Ad}_{g^{-1}} ( [ \lambda _{1} (x) , \lambda_{2} (x)]_{\mathfrak{g}})]^{\#}= X_{-[\lambda_{1},\lambda_{2}]_{\mathfrak{g}}},
\label{curvgtgt4}
\end{equation}
where the superscript $\#$ refers to a generalization of the fundamental vector field applicable to the case where the element of $\mathfrak{g}$ exhibits spacetime dependence\footnote{The sign on the right-hand side of \eqref{curvgtgt4} arises from using right group multiplication, as opposed to left multiplication \cite{Werner}.}. Remarkably, equation \eqref{curvgtgt4} reproduces the structure of equation \eqref{gtgt} (cf. \cite{wiena}).
%previously, $\Phi[\delta_{\text{GT}}(\lambda)] = X_{\lambda}$ and therefore, $X_{\lambda}$ is the vertical vector field related to the gauge symmetry. Now it's easy to see that the following commutator $[(\lambda_{1})^{\#}, X_{\lambda_{2}}]=0$ (the expected result for a left and right invariant vector field). So, in this case, it's obvious that the geometrical interpretation of the commutator \eqref{gtgt} is related with the right invariant vector fields in the fiber. 

We now analyze the Lie bracket of a vector field associated with a gauge transformation and one associated with an LT. Again, $\Omega (\tilde{\xi}, X_{\lambda}) = 0$.
Using this fact and equation \eqref{Cse}, we get
\begin{eqnarray}
\omega ([\tilde{\xi},X_{\lambda}]_{P})&=&\Omega (\tilde{\xi}, X_{\lambda})+\omega ([\tilde{\xi},X_{\lambda}]_{P})\nonumber\\
&=& \de_{P} \omega (\tilde{\xi}, X_{\lambda}) +\omega ([\tilde{\xi},X_{\lambda}]_{P})\nonumber\\
 &=& \tilde{\xi} [\text{Ad}_{g^{-1}} \lambda] - X_{\lambda} [\omega (\tilde{\xi})] \nonumber\\
 &=& \tilde{\xi} [\text{Ad}_{g^{-1}} \lambda] .
\label{curvaturehorvert1}
\end{eqnarray}
Therefore, we obtain \cite{mat3}
\begin{equation}
[\tilde{\xi} , X_{\lambda}]_{P} = (\tilde{\xi} [\text{Ad}_{g^{-1}} \lambda ])^{\#}=\left[\de_{P} [\text{Ad}_{g^{-1}} \lambda](\tilde{\xi})\right]^{\#},
\label{curvaturahorvert3}
\end{equation}
Using \eqref{diffeqlift1} and the fact that $\de_{P} \lambda (\tilde{X}) = \de \lambda (X)$, which follows from $\lambda(x) = \lambda (\pi (u))$, we can rewrite the right-hand side of \eqref{curvaturahorvert3} as
\begin{eqnarray}
 \de_{P} [\text{Ad}_{g^{-1}} \lambda] (\tilde{\xi}) 
 &=& \de_{P}[g^{-1} \lambda (x) g ] (\tilde{\xi}) \nonumber\\
& =& \de_{P} g^{-1} (\tilde{\xi} ) \lambda (x) g + g^{-1} \de_{P} \lambda (\tilde {\xi}) g + g^{-1} \lambda (x) \de_{P} g (\tilde{\xi}) \nonumber\\
&=& g^{-1} \mathcal{A} (\xi) \lambda (x) g - g^{-1} \lambda (x)\mathcal{A} (\xi) g + g^{-1} \de_{P} \lambda (\tilde{\xi}) g \nonumber\\
&=& g^{-1} ( \de \lambda (\xi) + [\mathcal{A} (\xi), \lambda (x) ]_{\mathfrak{g}})g \nonumber\\
&=&\text{Ad}_{ g^{-1}} \D \lambda (\xi).
 \label{eqliftgt}
\end{eqnarray}
Therefore,
\begin{equation}
[\tilde{\xi} , X_{\lambda}]_{P} = X_{\text{i}_\xi \D \lambda},
 \label{commutgttgt}
\end{equation}
which reproduces \eqref{ltgt}.

Recall that for a conventional fundamental vector field, $\lambda^{\#}$, that is associated with a spacetime-independent element of $\mathfrak{g}$, one obtains $[\tilde{\xi}, \lambda^{\#}]_{P} = 0$ \cite{Naka, Werner}. Therefore, to reproduce the symmetry algebra it is crucial to consider the spacetime dependence of the chosen algebra elements. Moreover, we can easily verify that our result elegantly reduces to the spacetime-independent case.
%More precisely, we observe that, from \eqref{Levhorizontal} and \eqref{eomglev}, we can rewrite
%\begin{equation}
 % \tilde{\xi} = R_{g *} \sigma_{*} \xi + [g^{-1} \de g (\xi)]^{\#} = R_{g *} \sigma_{*} \xi - [\text{Ad}_{g^{-1}} \mathcal{A}(\xi)]^{\#} = R_{g *} \sigma_{*} \xi - X_{\mathcal{A}(\xi)},
 % \label{levlift}
%\end{equation}
%where $\sigma$ is a local section of $P$. The fact that $[\tilde{\xi}, \lambda^{\#}]_{P} \neq 0$ arises directly the fact that we are effectively copying the Lie algebra across the fiber, and second, that $X_{\mathcal{A}(\xi)}$ and $X_{\lambda}$ are right-invariant vector fields.

Finally, we analyze the Lie bracket of two horizontal lifts. Note that the curvature on $P$ corresponds to the vertical component of the commutator of two horizontal vector fields (see \eqref{curvaturemean}). Using equations \eqref{Levhorizontal} and \eqref{eomglev} to express the horizontal lifts, we obtain
\begin{eqnarray}
[\tilde{\xi}_{1} , \tilde{\xi}_{2}]_{P} 
&=& [R_{g *} \sigma _{*} \xi_{1} - \{g^ {-1} \mathcal{A}(\xi _{1}) g\}^{\#},R_{g *} \sigma _{*} \xi _{2} - [g^{-1} \mathcal {A}(\xi_{2}) g]^{\#}]_{P}\nonumber \\
 &=& [R_{g *} \sigma _{*} \xi_{1} ,R_{g *} \sigma _{*} \xi_{2}]_{P} -[R_{g *} \sigma _{* } \xi_{1} ,\{g^{-1} \mathcal{A}(\xi_{2}) g\}^{\#}\} ]_{P}\nonumber\\
&&+[R_{g *} \sigma_ {*} \xi_{2},\{g^{-1} \mathcal{A}(\xi_{1}) g \}^{\#} ]_{P} +[\{g^{-1} \mathcal{A}(\xi_{1}) g\}^{\#},\{g^{-1} \mathcal{A }(\xi_{2}) g\}^{\#}]_{P} \nonumber\\
&=&R_{g *} \sigma _{*} [\xi _{1},\xi _{2}]_{M} -[R _{g *} \sigma _{*} \xi _{1} ,\{g^{-1 } \mathcal{A}(\xi _{2}) g\}^{\#} ]_{P}\nonumber\\
&& +[R_{g *} \sigma _{*} \xi _{2},\{g^{- 1} \mathcal{A}(\xi_{1}) g\}^{\#} ]_{P} - \{g^{-1}[ \mathcal{A}(\xi_{1}),\mathcal{A}(\xi_{2})]_{\mathfrak{g }} g\}^{\#},
\label{horccurvature}
\end{eqnarray} 
where we utilize
\begin{equation}
[R_{g *} \sigma _{*} \xi _{1} , \{g^{-1} \mathcal{A}(\xi _{2})
g\}^{\#}]_{P} = ( g^{-1} \xi_{1}[ \mathcal{A}(\xi_{2})] g )^{\#}.
\label{commutator1}
\end{equation}
and equations \eqref{curvgtgt2} and \eqref{curvgtgt4}. We can manipulate equation \eqref{horccurvature} further to yield:
\begin{eqnarray}
[\tilde{\xi}_{1}, \tilde{\xi}_{2}]_{P}
&=& R_{g *} \sigma_{*} [\xi_{1},\xi_{2 }]_M -(g^{-1} \xi_{1}[\mathcal{A}(\xi_{2})]g)^{\#}\nonumber\\
&&+(g^{-1} \xi_{2}[ \mathcal{A}(\xi_{1})]g)^{\#}
-\{g^{-1}[ \mathcal{A}(\xi_{1}),\mathcal{A}(\xi_{2})]_{\mathfrak{g}} g\}^{\#} \nonumber\\
&=& R_{g *} \sigma _{*} [\xi _{1},\xi _{2}]_M -\{g^{-1}( \xi _{1}[\mathcal{A}(\xi _{ 2})]- \xi_{2}[\mathcal{A}(\xi_{1})])g\}^{\#}
\nonumber\\
&& -\{g^{-1}[ \mathcal{A}(\xi_{1}),\mathcal{A}(\xi_{2})]_{\mathfrak{g}} g\}^{\#} \nonumber\\
%&=& R_{g *} \sigma_{*} [\xi_{1},\xi_{2}] -\{g^{-1}(\de \mathcal{A} (\xi_{1}, \xi_{2})+ \mathcal{A}[\xi_{1}, \xi_{2}])g\}^{\#}-\{g^{-1}[ \mathcal{A}(\xi_{1}),\mathcal{A}(\xi_{2})]_{\mathfrak{g}} g\}^{\#} \nonumber\\
&=& R_{g *} \sigma _{*} [\xi _{1},\xi _{2}]_M - \{g^{-1} \mathcal{A}[\xi _{1}, \xi _{2 }]_M)g\}^{\#}\nonumber\\
&&-\{g^{-1}(\de \mathcal{A} (\xi_{1}, \xi_{2}) +[ \mathcal{A}(\xi_{1}),\mathcal{A}(\xi_{2})]_{\mathfrak{g}} )g\}^{\#} \nonumber\\
%&=& \widetilde{[\xi_{1},\xi_{2}]}- [g^{-1} F(\xi_{1} , \xi_{2})g]^{\#} \nonumber\\
&=& \widetilde{[\xi_{1},\xi_{2}]_M}- [\Omega(\tilde{\xi}_{1},\tilde{\xi}_{2})]^{\#},
\label{conmlevh}
\end{eqnarray}
where, in the last line, we use \eqref{curvatureOF}. Once again, this result reproduces the commutator of two LTs, given in equation \eqref{ltlt}. Moreover, equation \eqref{conmlevh} is consistent with the geometric interpretation of the commutator of vector fields as the failure of the parallelogram formed by the flows associated with horizontal lifts to close \cite{Naka}. What is more, when projecting the right-hand side of equation \eqref{conmlevh} onto the base manifold, the gauge part vanishes and the remaining part coincides with the commutator on $M$, $[\xi_{1},\xi_{2}]_{M}$ (see also \cite{Prabhu_2017,Werner}).

In summary, we demonstrate that the symmetry algebra, as defined in equations \eqref{algebra}, admits a geometric interpretation as the Lie bracket of corresponding vector fields on $P$. A key observation underlying this result is the fact that, in $P$, there are no distinctions on the fields based on their transformation laws. This resolution accounts for the observed field-independence of the algebra. In the following section, we analyze the connection-independence of the presented construction.

\section{Connection independence}\label{othertransfs}

In the preceding section, we employ LTs to define the notion of horizontality. A natural question arises: do these results remain valid under alternative definitions of horizontality? Fortunately, any transformation that defines a notion of horizontality can be expressed as a particular combination of gauge transformations and diffeomorphisms (see \eqref{LTasDiffandGauge} for the corresponding expression for LTs). In this section, we extend our analysis to a generic connection and demonstrate that the symmetry algebra is invariant, reinforcing our geometric interpretation.

A generic combination of the relevant transformations can be written as $\delta_{\text{K}} (\xi) =\delta_{\text{Diff}}(\xi)- \delta_{\text{GT}}(\lambda[\xi])$, where $\lambda[\xi]\in \mathfrak{g}$ is an element of the Lie algebra that may dependent on the spacetime vector field $\xi$. Clearly, if gauge transformations and diffeomorphisms are symmetries of a given theory, then, $\delta_{\text{K}} (\xi)$ is also a symmetry. Moreover,
\begin{equation}
 \Phi[\delta_{\text{K}}(\xi)] 
 = \Phi[\delta_{\text{LT}}(\xi) + \delta_{\text{GT}} (\text{i}_{\xi} \mathcal{A}) - \delta_{\text{GT}}(\lambda[\xi]) ] = R_{g*} \sigma_{*} \xi - (g^{-1} \lambda[\xi] g)^{\#} ,
 %\tilde{\xi} + X_{\lambda[\xi]} + X_{A(\xi)},
 \label{PhiIn}
\end{equation}
where we use equations \eqref{LTasDiffandGauge} and \eqref{PhiSV}, and we apply $\Phi$ to each term separately. Importantly, equation \eqref{PhiIn} can be interpreted as stating that the horizontal lift associated with $\delta_{\text{K}}(\xi)$ is determined by a local connection, $\bar{\mathcal{A}}^{\mu}_{\ \nu}$, with the property that $\text{i}_\xi\bar{\mathcal{A}}^{\mu}_{\ \nu}=\lambda[\xi]^{\mu}_{\ \nu}$. 

Note that, for any two local connection fields, $\mathcal{A}^{\mu}_{\ \nu}$ and $\bar{\mathcal{A}}^{\mu}_{\ \nu}$, there exists a $1$-form valued in $\mathfrak{g}$, $K^{\mu}_{\ \nu}$, such that $\bar{\mathcal{A}}^{\mu}_{\ \nu}=\mathcal{A}^{\mu}_{\ \nu} + K^{\mu}_{\ \nu}$. Importantly, $K^{\mu}_{\ \nu}$ transforms \emph{covariantly} under gauge transformations. The local connection $\bar{\mathcal{A}}^{\mu}_{\ \nu}$ defines an associated covariant exterior derivative, $ \bar{\D}$. This derivative, when acting on a regular field $\Psi^{\mu}$, leads to 
\begin{equation}
 \bar{\D} \Psi^{\mu} = \de \Psi^{\mu} + \bar{\mathcal{A}}^{\mu}_{\ \nu} \wedge \Psi^{\nu}.
 \label{eAbar}
\end{equation}
Moreover, we can show that the curvature $2$-form associated with $\bar{\mathcal{A}}^{\mu}_{\ \nu}$, $ \bar{F}^{\mu}_{\ \nu} $, is related with that associated with $\mathcal{A}^{\mu}_{\ \nu}$ through
\begin{equation}
 \bar{F}^{\mu}_{\ \nu} = \de \bar{\mathcal{A}}^{\mu}_{\ \nu} + \bar{\mathcal{A}}^{\mu}_{\ \rho} \wedge \bar{\mathcal{A}}^{\rho}_{\ \nu} = F^{\mu}_{\ \nu} + \D K^{\mu}_{\ \nu} + K^{\mu}_{\ \rho} \wedge K^{\rho}_{\ \nu},
 \label{Curvaturerelations}
\end{equation}
where $F^{\mu}_{\ \nu} $
and $\D$ are, respectively, the curvature $2$-form and covariant exterior derivative associated with ${\mathcal{A}}^{\mu}_{\ \nu}$.

Equation \eqref{eAbar} allows us to write $\delta_{\text{K}} (\xi)$, acting on the regular field $\Psi^{\mu}$, as
\begin{equation}
 \delta_{\text{K}} (\xi) \Psi^{\mu} = [\delta_{\text{LT}} (\xi) - \delta_{\text{GT}} (\text{i}_\xi K)] \Psi^{\mu} = \bar{\text{D}} \text{i}_{\xi} \Psi^{\mu} + \text{i}_{\xi} \bar{\text{D}}\Psi^{\mu}.
 \label{KLT}
\end{equation}
Similarly, when we apply $\delta_{\text{K}} (\xi)$ on a connection field, it yields \cite{Prabhu_2017}
\begin{equation}
 \delta_{\text{K}}(\xi) \bar{\mathcal{A}}^{\mu}_{\ \nu} = \text{i}_{\xi}\bar{F}^{\mu}_{\ \nu}.
 \label{KLTcon1}
\end{equation}
This result follows from the transformation law of $\mathcal{A}^{\mu}_{\ \nu}$ and the expression that relate it with $ \bar{\mathcal{A}}^{\mu}_{\ \nu}$ and $K^{\mu}_{\ \nu}$.
%\begin{equation}
 % \delta_{\text{K}}(\xi) \mathcal{A}^{\mu}_{\ \nu} = \text{i}_{\xi}F^{\mu}_{\ \nu} - \D \text{i}_{\xi}K^{\mu}_{\ \nu} = \text{i}_{\xi}F^{\mu}_{\ \nu} - \D \text{i}_{\xi}K^{\mu}_{\ \nu} = \text{i}_{\xi}\bar{F}^{\mu}_{\ \nu} - \delta_{\text{K}}(\xi)K^{\mu}_{\ \nu}.
 % \label{KLTcon}
%\end{equation}
Importantly, equations \eqref{KLT} and \eqref{KLTcon1} links $\delta_{\text{K}}(\xi)$, which is the transformation used to give a notion of horizontality, with $\Psi^{\mu}$ and $\bar{\mathcal{A}}^{\mu}_{\ \nu}$.
%In other words, given a local connection, $\bar{\mathcal{A}}^{\mu}_{\ \nu}$, and it's $2$-form of curvature, we can find the transformation that satisfies \eqref{KLTcon1} and that provides us with another notion of horizontality in $P$ (whose connection projected to $M$ is $\bar{\mathcal{A}}^{\mu}_{\ \nu}$). 

We now proceed to compute the symmetry algebra involving $\delta_{\text{K}}(\xi)$. This calculation is straightforward due to the linearity of the commutator. It yields
\begin{eqnarray}
 [\delta_{\text{K}}(\xi_{1}),\delta_{\text{K}}(\xi_{2})]
 &=& [\delta_{\text{LT}}(\xi_{1}),\delta_{\text{LT}}(\xi_{2})] 
 -[\delta_{\text{LT}}(\xi_{1}),\delta_{\text{GT}}(\text{i}_{\xi_{2}}K)] \nonumber\\
 &&+[\delta_{\text{LT}}(\xi_{2}),\delta_{\text{GT}}(\text{i}_{\xi_{1}}K)]
 + [\delta_{\text{GT}}(\text{i}_{\xi_{1}}K),\delta_{\text{GT}}(\text{i}_{\xi_{2}}K)]\nonumber\\
 &=& \delta_{\text{K}}([\xi_{1},\xi_{2}]_{M}) - \delta_{\text{GT}}(\text{i}_{\xi_{2}} \text{i}_{\xi_{1}}\bar{F}),
 \label{KKTcomm2}
\end{eqnarray}
where we use equations \eqref{algebra} and \eqref{identidadpform}. Moreover,
%this commutator has the general structure obtained in \cite{Prabhu_2017}.
%\begin{equation}
%\begin{split}
 % [\delta_{\text{K}}(\xi_{1}),\delta_{\text{K}}(\xi_{2})= \delta_{\text{LT}}([\xi_{1},\xi_{2}]_{M}) &- \delta_{\text{GT}}(\text{i}_{\xi_{2}} \text{i}_{\xi_{1}}F)-\delta_{\text{GT}}(\text{i}_{[\xi_{1},\xi_{2}]_{M}}K) \\ &-\delta_{\text{GT}}(\text{i}_{\xi_{2}} \text{i}_{\xi_{1}} \D K)-\delta_{\text{GT}}[\text{i}_{\xi_{2}} \text{i}_{\xi_{1}} (K\wedge K)],
 %\label{KKTcomm1}
%\end{split}
%\end{equation}
equation \eqref{KKTcomm2} is a direct generalization of equation \eqref{ltlt}: the commutator produces a $\delta_{\text{K}}$ transformation whose generator is $[\xi_{1},\xi_{2}]_{M}$ plus the gauge transformation related with the curvature associated of the connection $\bar{\mathcal{A}}^{\mu}_{\ \nu}$. Analogously, we obtain
\begin{equation}
 [\delta_{\text{K}} (\xi),\delta_{\text{GT}} (\lambda)] = \delta_{\text{GT}} (\text{i}_{\xi} \bar{\D} \lambda).
 \label{KGT}
 \end{equation}
Comparing with \eqref{ltgt}, we observe that it is of the same form. Moreover, since the gauge commutator is independent of the local connection, we conclude that the algebra \eqref{algebra} retains its structure regardless of the connection we use.

Therefore, the symmetry algebra ---and its geometric interpretation--- are independent of the notion of horizontality. To conclude this section, we present an example where we use a specific local connection.

\subsection{Lorentz-Lie transformation and the Jacobson-Mohd conundrum}

Recently, a particular combination of gauge transformation and diffeomorphisms, known as the Lorentz-Lie transformation, has been studied in the context of gravity theories in the first-order formalism \cite{Jacobson}. In this context, the dynamical fields consist of the tetrad $1$-form $e^{\mu}$ and a ``spin connection'' $1$-form $\omega^{\mu}_{\ \nu}$. Following \cite{Jacobson}, we adopt the assumption of vanishing torsion ($\D e^{\mu} = 0$). This implies that the spin connection is completely determined by the tetrads. Moreover, as the term ``tetrad'' suggests, spacetime has four dimensions.  Also, the gauge group is the Lorentz group $SO(1,3)$, with its associated Lie algebra $\mathfrak{so}(1,3)$.

We argue that the Lorentz-Lie transformations are particularly useful for the authors of \cite{Jacobson} for its properties when associated with Killing vector fields, which encode spacetime symmetries \cite[Appendix C]{WaldB}. Let $\xi$ be a Killing vector field, then, by definition
\begin{equation}\label{Killing}
\mathcal{L}_{\xi} g = 0,
\end{equation}
where $g$ is the metric tensor. Equation \eqref{Killing} is equivalent to $\nabla^\mu \xi^\nu + \nabla^\nu \xi^\mu = 0$, where $\nabla_\mu$ denotes the metric-compatible and torsionless covariant derivative along the vector field dual to $e^{\mu}$. Also, $\xi^\mu = e^\mu(\xi)$, and we use the components of the inverse metric (metric) in the tetrad (dual tetrad) basis, denoted by $\eta^{\mu\nu}$ ($\eta_{\mu\nu}$), to raise (lower) Greek indices; $\eta^{\mu\nu}$ ($\eta_{\mu\nu}$) has the matricial form of the Minkowski metric. In terms of the tetrad, equation \eqref{Killing} becomes
\begin{equation}
\mathcal{L}_{\xi} e^{\mu} = -(\lambda_{\xi})^{\mu}_{\ \nu} e^{\nu},
\label{Killingviel}
\end{equation}
where $(\lambda_{\xi})^{\mu\nu} + (\lambda_{\xi})^{\nu\mu} = 0$, making the right-hand side an gauge transformation. Furthermore, using a well-known relation involving the second covariant derivative of a Killing vector field, we find that, for the spin connection,
\begin{equation}
\mathcal{L}_{\xi} \omega^{\mu}_{\ \nu} = \D (\lambda_{\xi})^{\mu}_{\ \nu},
\label{KillingConnection}
\end{equation}
which is also a gauge transformation.

As always, there exists a $1$-form $K^{\mu}_{\ \nu}$ such that the connection associated with the Lorentz-Lie transformation can be expressed as $ \bar{\omega}^{\mu}_{\ \nu} = \omega^{\mu}_{\ \nu} + K^{\mu}_{\ \nu}$. We define the Lorentz-Lie transformation along an arbitrary vector field $\xi$ by requiring $K^{\mu\nu}$ to be such that
\begin{equation}
\text{i}_{\xi} K^{\mu\nu} = \frac{1}{2} \left( \nabla^{\mu} \xi^{\nu} - \nabla^{\nu} \xi^{\mu} \right).
\label{ConnnectionKoss}
\end{equation}
Interestingly, equation \eqref{ConnnectionKoss} has been identified elsewhere with the ``momentum map'' \cite{2018arXiv180205345B,Elgood:2020svt}.
%For a Lorentz-Lie transformation along $\xi$, denoted as $\delta_{\text{LL}}(\xi)$, we obtain a transformation analogous to equation \eqref{Killingviel}, but where $(\lambda_{\xi})^{\mu\nu} = \nabla^{\mu} \xi^{\nu} + \text{i}_{\xi}\omega^{\mu\nu}$. Note that for $(\lambda_{\xi})^{\mu\nu}$ to be antisymmetric in its algebra indexes, we require $\xi$ to be a Killing vector field (recall that $\omega^{\mu\nu} + \omega^{\nu\mu} = 0$).

Let $\delta_{\text{LL}}(\xi)$ denote the Lorentz-Lie transformation along $\xi$. Interestingly, when $\xi$ is a Killing vector field
\begin{equation}
 \delta_{\text{LL}}(\xi) e^{\mu}= \frac{1}{2} \left( \nabla^{\mu} \xi_{\nu} + \nabla_{\nu} \xi^{\mu} \right) e^{\nu}=0.
 \label{LorentzLieKilling}
\end{equation}
Analogously, $\delta_{\text{LL}}(\xi) \omega^{\mu}_{\ \nu} = 0$, as expected from the torsionless condition. Thus, the Lorentz-Lie transformations along Killing vector fields leaves the dynamical fields invariant, as noted in \cite{Jacobson}.
%This is the main reason the authors in \cite{Jacobson} find this transformation useful when calculating the thermodynamical properties of stationary black holes. 
%
In addition, the curvature $2$-form associated with $\bar{\omega}^{\mu}_{\ \nu}$, $\bar{F}^{\mu \nu}$, vanishes when contracted with a Killing vector field. Contracting \eqref{Curvaturerelations} with $\xi$ and $\chi$ leads to
\begin{equation}
 \text{i}_{\chi} \text{i}_{\xi}\bar{F}^{\mu \nu} = \text{i}_{\chi} \text{i}_{\xi}R^{\mu \nu} + \text{i}_{\chi} \text{i}_{\xi}\D K^{\mu \nu} + \text{i}_{\xi}K^{\mu}_{\ \alpha} \text{i}_{\chi} K^{\alpha \nu} - \text{i}_{\chi} K^{\mu}_{\ \alpha} \text{i}_{\xi} K^{\alpha \nu},
 \label{LLcurvature}
\end{equation}
where $R^{\mu \nu}$ is the curvature 2-form associated with $\omega^{\mu \nu}$, as is conventionally denoted. We can manipulate the second term on the right-hand side using \eqref{identidad2tensor} to obtain
\begin{eqnarray}
 \text{i}_{\chi} \text{i}_{\xi}\D K^{\mu \nu} &=& \text{i}_{\xi} \D \text{i}_{\chi} K^{\mu \nu}
- \text{i}_{\chi} \D \text{i}_{\xi} K^{\mu \nu} -\text{i}_{[\xi,\chi]} K^{\mu \nu} \nonumber \\
&=& \frac{1}{2} \left( \xi^{\alpha} \nabla_{\alpha} [\nabla^{\mu} \chi^{\nu} - \nabla^{\nu} \chi^{\mu}] - \chi^{\alpha} \nabla_{\alpha} [\nabla^{\mu} \xi^{\nu} - \nabla^{\nu} \xi^{\mu}] \right) \nonumber \\
&& -\frac{1}{2} \left( \nabla^{\mu}[\xi^{\alpha} \nabla_{\alpha} \chi^{\nu} - \chi^{\alpha} \nabla_{\alpha} \xi^{\nu}] - \nabla^{\nu}[\xi^{\alpha} \nabla_{\alpha} \chi^{\mu} - \chi^{\alpha} \nabla_{\alpha} \xi^{\mu}]\right) \nonumber \\
&=& -\text{i}_{\chi} \text{i}_{\xi}R^{\mu \nu} -\frac{1}{2} \left( \nabla^{\mu}\xi^{\alpha} \nabla_{\alpha} \chi^{\nu} - \nabla^{\mu}\chi^{\alpha} \nabla_{\alpha} \xi^{\nu} \right) \nonumber \\
&& + \frac{1}{2} \left( \nabla^{\nu}\xi^{\alpha} \nabla_{\alpha} \chi^{\mu} - \nabla^{\nu}\chi^{\alpha} \nabla_{\alpha} \xi^{\mu}\right),
\label{LL2formsaturated}
\end{eqnarray}
where we use the Bianchi identities, the Riemann tensor definition in terms of $\nabla_\mu$, and the relation of such a tensor with $R^{\mu\nu}$. Inserting expression \eqref{LL2formsaturated} into \eqref{LLcurvature} yields
\begin{eqnarray}
 \text{i}_{\chi} \text{i}_{\xi}\bar{F}^{\mu \nu} &=& -\frac{1}{2} \left( \nabla^{\mu}\xi^{\alpha} \nabla_{\alpha} \chi^{\nu} - \nabla^{\mu}\chi^{\alpha} \nabla_{\alpha} \xi^{\nu} \right) \nonumber\\
 &&+ \frac{1}{2} \left( \nabla^{\nu}\xi^{\alpha} \nabla_{\alpha} \chi^{\mu} - \nabla^{\nu}\chi^{\alpha} \nabla_{\alpha} \xi^{\mu}\right) \nonumber \\
 &&+ \frac{1}{4} \left( \nabla^{\mu}\xi^{\alpha} \nabla_{\alpha} \chi^{\nu}-\nabla^{\alpha}\xi^{\mu} \nabla_{\alpha} \chi^{\nu} - \nabla^{\mu}\xi^{\alpha} \nabla^{\nu} \chi_{\alpha}+\nabla^{\alpha}\xi^{\mu} \nabla^{\nu} \chi_{\alpha} \right) \nonumber \\ &&- \frac{1}{4} \left( \nabla^{\mu}\chi^{\alpha} \nabla_{\alpha} \xi^{\nu}-\nabla^{\alpha}\chi^{\mu} \nabla_{\alpha} \xi^{\nu} - \nabla^{\mu}\chi^{\alpha} \nabla^{\nu} \xi_{\alpha}+\nabla^{\alpha}\chi^{\mu} \nabla^{\nu} \xi_{\alpha} \right) \nonumber \\
 &=&\frac{1}{4} \left([\nabla^{\mu}\chi^{\alpha} + \nabla^{\alpha}\chi^{\mu}] [\nabla_{\alpha} \xi^{\nu}+\nabla^{\nu} \xi_{\alpha}] \right) \nonumber\\
 &&-\frac{1}{4} \left([\nabla^{\mu}\xi^{\alpha} + \nabla^{\alpha}\xi^{\mu}] [\nabla_{\alpha} \chi^{\nu}+\nabla^{\nu} \chi_{\alpha}] \right).
 \label{curvatureLL}
\end{eqnarray}
Hence, when $\xi$ and/or $\chi$ are Killing vector fields, $ \text{i}_{\chi} \text{i}_{\xi} \bar{F}^{\mu\nu} = 0$. 

Inspecting \eqref{KKTcomm2}, we see that the commutator of two Lorentz-Lie transformations, when one of these transformations is along a Killing field, reduces to a Lorentz-Lie transformation associated with $[\xi, \chi]_{M}$. In other words, this commutator has no additional ``gauge term'' (cf. \cite{Ortin:2002qb,Fatibene2011,Elgood:2020svt}), which simplifies the computation of black hole thermodynamics.  

The fact that Lorentz-Lie transformations along Killing fields leave the dynamical fields invariant, together with the absence of an additional gauge term in their commutator, may explain their effectiveness in calculating black hole thermodynamics \cite{Jacobson}. Notably, the absence of a gauge term in this commutator reproduces the result of \cite{wald}, as argued in Refs. \cite{Simone,Prabhu_2017,Jacobson,toolkit,Elgood2020mdx}. However, as evident from equations \eqref{Killingviel} and \eqref{KillingConnection}, a general local connection inevitably introduces a gauge contribution to the commutator of the corresponding transformations. This contribution, which is absent in the metric formalism, should nevertheless be manageable.

\section{Conclusions}\label{conc}

In this work, we use the first-order formalism to study gauge and gravity theories. We analyze the structure of symmetries and demonstrate that their commutators form a closed algebra. Remarkably, we also show that these commutators are independent of whether they are computed by acting on regular fields or on connection fields, despite the distinct nature of these calculations. This observation points to an underlying geometric structure.

The geometric structure we identify is associated with principal fiber bundles, where all the fields have the same transformation properties. Specifically, we construct vector fields on the fiber bundle that encode the relevant symmetries. The key structure is provided by the connection on the principal fiber bundle, which selects vertical and horizontal directions. Vertical directions correspond to vectors associated with the fibers, which are isomorphic to the Lie group, and thus represent gauge transformations, while horizontal directions are associated with, say, local translations (LTs). Importantly, the Lie bracket of the vector fields representing the symmetries reproduces the algebra originally derived by acting on the dynamical fields, uncovering the geometric structure behind the symmetry algebra.

The formalism we develop is fundamentally kinematical, with the action only playing the role of defining the theories' symmetries: gauge and diffeomorphism invariance. In fact, these symmetries are present simply because there are no nondynamical fields in the actions \cite{BonCris,YuriCristobal2}. Extending our framework to include nondynamical fields, which would ``break'' the symmetries, represents a promising avenue for future research. Such generalizations could offer new mathematical insights and broaden the applicability of our approach.

Another intriguing avenue of research involves studying principal fiber bundles with nontrivial topology. These arise when the base manifold, $M$ (which in our case represents spacetime), exhibits a nontrivial topology, leading to fiber bundles that are not globally $M \times G$. We anticipate that these scenarios could unlock new applications of our formalism, particularly in gravitational theories. For example, these topological features could facilitate investigations of torsion effects. A compelling direction is the potential generalization of the Aharonov-Bohm effect to gravity theories, framed within principal fiber bundles. Such a generalization could pave the way for experimental proposals that reduce noise, improving upon earlier methods, such as the one described in Ref. \cite{YuriBruno}.

Additionally, the study of symmetries of the theories in asymptotically flat spacetimes, where one can define the ADM mass and angular momentum \cite{Arnowitt}, is particularly interesting. For spacetimes that are also stationary, our formalism could offer new methods to compute Noether charges and produce alternative derivations of the first law of black hole mechanics \cite{wald}. Such tools could provide a deeper geometric interpretation of these laws through the lens of principal fiber bundle geometry, offering fresh perspectives on this topic.

\ack
We thank C. Chryssomalakos and C. Corral for their insightful feedback and constructive suggestions. This project was supported by CONAHCYT through the FORDECYT-PRONACES grant No. 140630 and the UNAM DGAPA-PAPIIT grant No. IN101724, as well as by CONAHCYT’s Graduate Scholarship Program.

\appendix

\section{Basic definitions}\label{defs}

This Appendix draws on reference \cite{Naka} and the Appendix of \cite{review}. It provides a brief overview of the basic definitions of differential forms; for specific conventions, we refer the reader to \cite{Naka}.

A $p$-form is a totally antisymmetric $(0,p)$ tensor. Naturally, for a $p$-form to be nontrivial, it must satisfy $p \leq N$, where $N$ is the manifold dimensionality. Differential forms can be multiplied using the wedge product, $\wedge$, which is essentially an antisymmetrized tensor product. This operation takes a $q$-form and an $r$-form as input and produces a $(q+r)$-form as output.

The exterior derivative, $\de$, is a map from $p$-forms to $p+1$-forms defined by acting with any torsionless derivative on the $p$-form and antisymmetrizing all the spacetime indexes. Another operator we use is the inner derivative along the vector field $\xi$, $\text{i}_\xi$, which acts on a $p$-form and yields a $p-1$-form. This $p-1$-form is obtained by saturating the original form with $\xi$. Both operators follow a graded Leibniz rule, namelt, if $\omega$ is a $q$-form, then
 \begin{eqnarray}
 \de(\omega \wedge \eta) &=& \de \omega \wedge \eta + (-1)^{q} \omega \wedge \de \eta, \label{Leibniz1}\\
 \text{i}_{\xi}(\omega \wedge \eta) &=& \text{i}_{\xi} \omega \wedge \eta + (-1)^{q} \omega \wedge \text{i}_{\xi} \eta.\label{Leibniz2}
 \end{eqnarray}
Remarkably, the Lie derivative of a $q$-form $\omega$ along $\xi$ can be written as
 \begin{equation}
 \mathcal{L}_{\xi} \omega = (\de \text{i}_{\xi} + \text{i}_{\xi} \de) \omega.
 \label{magica}
 \end{equation}
This is the celebrated Cartan's magic formula.

We can also define a covariant exterior derivative, $\D$, which extends the concept of the exterior derivative to $\mathfrak{g}$-valued $p$-forms, such as $T^{\mu_1 \dots \mu_l}{}_{\nu_1 \dots \nu_k}$. Unlike the standard exterior derivative, $\D$ ensures that the $p$-form transforms covariantly under gauge transformations. It is defined as follows:
 \begin{eqnarray}
 \D{} T^{\mu_1\dots \mu_l}{}_{{\nu_1\dots \nu_k}} &=& \de{} T^{\mu_1\dots \mu_l}{}_{{\nu_1\dots \nu_k}} + \mathcal{A}^{\mu_1}_{\ \rho}T^{\rho\dots \mu_l}{}_{{\nu_1\dots \nu_k}} +\dots +\mathcal{A}^{\mu_l}_{\ \rho}T^{\mu_1\dots \rho}{}_{{\nu_1\dots \nu_k}}\nonumber\\
 && - \mathcal{A}^{\rho}_{\ \nu_1}T^{\mu_1\dots \mu_l}{}_{{\rho\dots \nu_k}} -\dots -\mathcal{A}^{\rho}_{\ \nu_k}T^{\mu_1\dots \mu_l}{}_{{\nu_1\dots \rho}}.
 \label{def D}
 \end{eqnarray}
We can show that $\D$ satisfies a graded Leibniz rule analogous to \eqref{Leibniz1}.

Moreover, any connection $1$-form $\mathcal{A}^{\mu}_{\ \nu}$ defines the following curvature $2$-form:
 \begin{equation}\label{def curvature}
 F^{\mu}_{\ \nu} = \de \mathcal{A}^{\mu}_{\ \nu} + \mathcal{A}^{\mu}_{\ \rho}\wedge \mathcal{A}^{\rho}_{\ \nu}.
 \end{equation}
We can easily show that this curvature $2$-form satisfies the Bianchi identity:
 \begin{equation}\label{Bianchi}
\D{} F^{\mu}_{\ \nu} =0.
 \end{equation}
In addition, we can verify that
 \begin{equation}\label{Dsquared}
\D{}\D{} \Psi^{\mu} = F^{\mu}_{\ \nu}\wedge \Psi^{\nu}.
 \end{equation}
This generalizes in a straightforward manner to forms with additional algebra indexes: for each index, a corresponding term is generated, with the condition that indexes in the lower position have a minus sign.

Finally, we note that any two vector fields in a manifold define a third vector field via its Lie bracket. If we consider spacetime $M$ as the manifold and an arbitrary function $f$ on it, this bracket, for the $\xi_1$ and $\xi_2$ vector fields, takes the form
\begin{equation}
 [\xi_1,\xi_2]_M (f) = \xi_1 [\xi_2 (f)] - \xi_2 [\xi_1 (f)]. \label{LieBracket}
\end{equation}

\section{Identities}\label{identities}

In this appendix, we deduce some identities that are relevant throughout the paper. The starting point is $[\mathcal{L}_{\xi_{1}}, \text{i}_{\xi_{2}} ] = \text{i}_{[\xi_{1},\xi_{2}]_{M}}$ \cite{Gockeler:1987an}. Acting on an algebra-valued $p $-form, $\phi^{\mu}$, we get
\begin{eqnarray}
\text{i}_{[\xi_{1},\xi_{2}]_{M}} \phi^{\mu} 
&=& \mathcal{L}_{\xi_{1}} \text{i}_{\xi_{2}} \phi^{\mu} - \text{i}_{\xi_{2}} \mathcal{L}_{\xi_{1}} \phi^{\mu} \nonumber\\
&=& \de \text{i}_{\xi_{1}} \text{i}_{\xi_{2}} \phi^{\mu} + \text{i}_{\xi_{1}} \de \text{i}_{\xi_{2}} \phi^{\mu}- \text{i}_{\xi_{2}}\de \text{i}_{\xi_{1}} \phi^{\mu} - \text{i}_{\xi_{2}} \text{i}_{\xi_{1}} \de \phi^{\mu} \nonumber\\
&= &\D \text{i}_{\xi_{1}} \text{i}_{\xi_{2}} \phi^{\mu} - \mathcal{A}^{\mu}_{\ \nu} \text{i}_{\xi_{1}} \text{i}_{\xi_{2}} \phi^{\nu} + \text{i}_{\xi_{1}} (\D \text{i}_{\xi_{2}} \phi^{\mu}-\mathcal{A}^{\mu}_{\ \nu} \text{i}_{\xi_{2}} \phi^{\nu} )\nonumber\\ 
 & & - \text{i}_{\xi_{2}}(\D \text{i}_{\xi_{1}} \phi^{\mu}-\mathcal{A}^{\mu}_{\ \nu} \text{i}_{\xi_{1}} \phi^{\nu} ) - \text{i}_{\xi_{2}} \text{i}_{\xi_{1}} (\D \phi^{\mu} - \mathcal{A}^{\mu}_{\ \nu} \phi^{\nu}) \nonumber\\
 &=& \D \text{i}_{\xi_{1}} \text{i}_{\xi_{2}} \phi^{\mu} + \text{i}_{\xi_{1}} \D \text{i}_{\xi_{2}} \phi^{\mu}- \text{i}_{\xi_{2}}\D \text{i}_{\xi_{1}} \phi^{\mu} - \text{i}_{\xi_{2}} \text{i}_{\xi_{1}} \D \phi^{\mu}.
\label{identidadpform}
\end{eqnarray}
Furthermore, we can easily verify that the identity \eqref{identidadpform} holds when we apply it to tensors with additional algebra indexes. In particular, for a $p$-forms of the form $T^{\mu}_{\ \nu}$, it becomes
\begin{equation}
\text{i}_{[\xi_{1},\xi_{2}]_{M}} T^{\mu}_{\ \nu} = \D \text{i}_{\xi_{1}} \text{i}_{\xi_{2}} T^{\mu}_{\ \nu} + \text{i}_{\xi_{1}} \D \text{i}_{\xi_{2}}T^{\mu}_{\ \nu} - \text{i}_{\xi_{2}}\D \text{i}_{\xi_{1}}T^{\mu}_{\ \nu} - \text{i}_{\xi_{2}} \text{i}_{\xi_{1}} \D T^{\mu}_{\ \nu} .
\label{identidad2tensor}
\end{equation}

Using equation \eqref{identidadpform}, we can manipulate $\text{i}_{[\xi_{1}, \xi_{2}]_{M}} \D \Psi^{\mu}$ and $\D \text{i}_{[\xi_{1}, \xi_{2}]_{M}} \Psi^{\mu}$, obtaining,
\begin{eqnarray}
\text{i}_{[\xi_{1}, \xi_{2}]_{M}} \D \Psi^{\mu}
&=&\D \text{i}_{\xi_{1}} \text{i}_{\xi_{2}} \D \Psi^{\mu} + \text{i}_{\xi_{1}} \D \text{i}_{\xi_{2}} \D \Psi^{\mu}- \text{i}_{\xi_{2}}\D \text{i}_{\xi_{1}} \D \Psi^{\mu} - \text{i}_{\xi_{2}} \text{i}_{\xi_{1}} \D \D \Psi^{\mu}\nonumber \\
&=& \D \text{i}_{\xi_{1}} \text{i}_{\xi_{2}} \D \Psi^{\mu} + \text{i}_{\xi_{1}} \D \text{i}_{\xi_{2}} \D \Psi^{\mu}- \text{i}_{\xi_{2}}\D \text{i}_{\xi_{1}} \D \Psi^{\mu}-\text{i}_{\xi_{2}} \D \D \text{i}_{\xi_{1}} \Psi^{\mu} \nonumber\\
& & - \text{i}_{\xi_{2}}\text{i}_{\xi_{1}} F^{\mu}_{\ \nu} \Psi^{\nu}+ \text{i}_{\xi_{1}} F^{\mu}_{\ \nu} \wedge \text{i}_{\xi_{2}} \Psi^{\nu},
\label{iden5}\\
\D \text{i}_{[\xi_{1}, \xi_{2}]_{M}} \Psi^{\mu}
&=&\D \D \text{i}_{\xi_{1}} \text{i}_{\xi_{2}} \Psi^{\mu} + \D \text{i}_{\xi_{1}} \D \text{i}_{\xi_{2}} \Psi^{\mu}- \D \text{i}_{\xi_{2}}\D \text{i}_{\xi_{1}} \Psi^{\mu} - \D \text{i}_{\xi_{2}} \text{i}_{\xi_{1}} \D \Psi^{\mu} \nonumber\\
&=& \text{i}_{\xi_{1}} \D \D \text{i}_{\xi_{2}} \Psi^{\mu} - \text{i}_{\xi_{1}} F^{\mu}_{\ \nu} \wedge \text{i}_{\xi_{2}} \Psi^{\nu}+ \D \text{i}_{\xi_{1}} \D \text{i}_{\xi_{2}} \Psi^{\mu}\nonumber\\
&&- \D \text{i}_{\xi_{2}}\D \text{i}_{\xi_{1}} \Psi^{\mu} - \D \text{i}_{\xi_{2}} \text{i}_{\xi_{1}} \D \Psi^{\mu} .
\label{iden6}
\end{eqnarray}
We use these expressions in the main body of the paper.

\section{Principal fiber bundles}\label{PFP}

A principal fiber bundle over $M$ with fiber $G$ is a mathematical structure consisting of a total space, $P$, a base space, $M$, a Lie group $G$ (the ``fiber''), and a projection map $\pi: P \to M$, such that $P$ is a smooth manifold that locally looks like $M \times G$. We can define a connection $1$-form on $P$, $\omega$, that has the information on how $T_{u}P$, the tangent space over $u \in P$, splits into the vertical and horizontal parts, which we denote, respectively, by $V_{u}P$ and $H_{u}P$. Importantly, the manifolds associated with the vertical directions are isomorphic to $G$.

We say that $X$ is a vertical vector field if $\pi_{*} X =0$. To construct elements in $V_{u}P$, we can take $\lambda \in \mathfrak{g}$ and define a curve in $P$, by acting with the group right multiplication, $R_g$, as
\begin{equation}
R_{ \exp(t \lambda)} u = u \exp(t \lambda),
\label{vercur}
\end{equation}
where $t$ is the parameter of the curve. Note that $\pi (u) = \pi (u \exp(t \lambda)) $, and therefore \eqref{vercur} is on a single fiber for all $t$. We define $\lambda^{\#} \in V_{u}P $ as the tangent vector to the curve \eqref{vercur} at $t=0$. Concretely, $\lambda^{\#}  $ acts on $f : P \rightarrow \mathbb{R}$ as
\begin{equation}
\lambda^{\#} f(u) = \frac{d}{dt} f (u \exp(t \lambda)) \bigg{|}_{t=0}.
\label{vup}
\end{equation}
If we repeat the previous construction on each fiber, we obtain the vector field $\lambda^{\#}$, which is known as the fundamental vector field generated by $\lambda$. In fact, the map $\#$ is an isomorphism between $\mathfrak{g}$ and $V_{u}P$ \cite{Naka}.

We define the connection 1-form, $\omega$, also known as the Ehresmann connection or simply the connection in $P$, by 
\begin{equation}
\omega (\lambda^{\#}) = \lambda,
\label{prop1}
\end{equation}
and
\begin{equation}
R_{g}^{*} \omega = g^{-1} \omega g,
\label{prop2}
\end{equation}
for all $\lambda \in \mathfrak{g}$ and $g \in G$. Importantly, $\omega$ enables us to characterize $H_{u}P$ by the condition that any horizontal vector $X$ satisfies
\begin{equation}
\omega(X) = 0.
\label{hup}
\end{equation}

Consider a collection of open sets $\{U_{i}\}$ covering $M$. A local section is a smooth map $\sigma_i: U_i \to P$ such that $\pi \circ \sigma_i $ is the identity map on $U_i $. This means that $\sigma_i(x)$ is a point in $P$ ``on top of'' $x\in M$. We define the $1$-form valued at $\mathfrak{g}$, $\mathcal{A}_{i} = \sigma_{i}^{*} \omega$ in $U_{i}$, which we call the local connection and which corresponds, in the context of gauge theories, to a gauge field (the algebra indexes are omitted for simplicity). On the other hand, given a section $\sigma$ and a local connection $\mathcal{A} = \sigma^{*} \omega$ in $M$, there exists $\omega$, a $1$-form connection on $P$, such that $\sigma^{*} \omega = \mathcal{A}$; in fact, we can express $\omega$ as
\begin{equation}
\omega = g^{-1} \pi^{*} \mathcal{A} g + g^{-1} \text{d}_{P} g,
\label{connMP}
\end{equation}
where $\text{d}_{P}$ is the exterior derivative in $P$. We can show that this $\omega$ satisfies \eqref{prop1} and \eqref{prop2} \cite{Naka}. Another relevant property of $\omega$ is that, by construction, it is global in $P$ \cite{Naka}, that is, given two open sets $U_{i}$ and $U_{j}$, $\omega |_{U_{i}} = \omega |_{U_{j}}$ in $U_{i} \cap U_{j}$.
This relation generates the transformation property of the local connection.

Consider a curve, $\gamma$, in $M$. We say that a curve $\tilde{\gamma} \subset P$ is the horizontal lift of $\gamma$ if $\pi \circ \tilde{\gamma} = \gamma$ and if the tangent to $\tilde{\gamma}$, denoted $\tilde{X}$, is a horizontal vector. hence, $\omega(\tilde{X})=0$, which is a differential equation, ensuring that $\tilde{\gamma}$ exists and it is unique. We can express $\tilde{X}$ as a function of $X$, the tangent vector to $\gamma$. To do this, we consider a local section $\sigma$ that allows us to write $\tilde{\gamma} (t)= \sigma(\gamma(t)) g(t)$ where $g(t) \in G$. Then, the tangent vector can be expressed as $\tilde{X} = \tilde{\gamma}_{*} X$, which we can, in turn, rewrite as \cite{Naka}
\begin{equation}
\tilde{X}=R_{g(t) *}( \sigma _{*} X) + [g(t)^{-1} \text{d} g (X)]^{\#}.
\label{Levhorizontal}
\end{equation}
Notice that, at each $u\in P$, $\pi_{*}\tilde{\xi} = \xi$. Also, from equation \eqref{Levhorizontal} and $\omega(\tilde{X})=0$, we can derive a differential equation for the element $g$ that generates the horizontal lifting of the curve. This equation takes the form
\begin{equation}
\text{d}g(X) = \frac{d g(t)}{dt} = - \mathcal{A}(X) g(t).
\label{eomglev}
\end{equation}
We can apply this construction to relate vector fields on $M$ and horizontal lifts. This relationship is unique and the resulting vectors satisfy $\pi_{*} \tilde{X} = X$ \cite{Werner}.

We can also define a curvature, $\Omega$, associated with the connection $1$-form $\omega$ using Cartan's structure equation. The definition is
\begin{equation}
\Omega = \text{d}_{P} \omega + \omega \wedge \omega.
\label{Cse}
\end{equation}
The geometric interpretation of $ \Omega$ arises when we apply  it to $X,Y\in H_{u}P$:
\begin{equation}\label{curvaturemean}
\Omega (X,Y) = \text{d}_{P} \omega (X,Y) = X[\omega(Y)] - Y[\omega(X)] - \omega([X,Y ]_P) = - \omega([X,Y]_P),
\end{equation}
where $[\cdot,\cdot]_P$ is the Lie bracket in $P$. Thus, the horizontal lifting of a closed infinitesimal parallelogram in $ M $, associated with two vectors whose lifts correspond to the fields $ X $ and $ Y $, describes the failure of the lifted circuit to close. This failure is vertical and is captured by $ \Omega(X,Y) $.

Using equation \eqref{connMP}, we can calculate $\Omega$. The first term of $\Omega$ produces
\begin{eqnarray}
\de_{P} \omega &=& \de_{P}g^{-1} \pi^{*} \mathcal{A} g +g^{-1} \de_{p} \pi^{*} \mathcal{A}g- g^{-1} \pi^{*} \mathcal{A} \wedge \de_{p} g +\de_{p} g^{-1} \wedge \text{d }_{P} g \nonumber\\
&=& -(g^{-1} \de_{p} g g^{-1} ) \wedge \pi^{*} \mathcal{A} g +g^{-1} \pi^{*} \ of \mathcal{A}g\nonumber\\
&&- g^{-1} \pi^{*} \mathcal{A} \wedge \de_{p} g -(g^{-1} \de_{p} g ) \wedge g^{-1} \text{d}_{P} g.
\label{CSEcurv}
\end{eqnarray}
On the other hand, for the second term of $\Omega$ we have
\begin{eqnarray}
\omega \wedge \omega &= &(g^{-1} \pi^{*} \mathcal{A} g +g^{-1} \de_{p}g) \wedge (g^{-1} \pi^{*} \mathcal{A} g +g^{-1} \de_{p}g) \nonumber\\
&=&g^{-1} \pi^{*} (\mathcal{A} \wedge \mathcal{A} )g +g^{-1} \pi^{*} \mathcal{A} \wedge \de_{P} g \nonumber\\
&&+(g^{-1} \de_{p}g g^{-1}) \wedge \pi^{*} \mathcal{A} g +(g^{-1} \de_{ p}g ) \wedge (g^{-1} \de_{p}g ).
\label{wwedw}
\end{eqnarray}
Adding \eqref{CSEcurv} and \eqref{wwedw}, we find
\begin{equation}
\Omega = g^{-1} \pi^{*}( \de \mathcal{A} + \mathcal{A} \wedge \mathcal{A}) g = g^{-1} \pi^{* } F g,
\label{curv2E}
\end{equation}
where $F^{\mu}_{\ \nu}=\de \mathcal{A}^{\mu}_{\ \nu}+ \mathcal{A}^{\mu}_{\ \rho} \wedge \mathcal{A}^{\rho}_{\ \nu}$ is called the field strength of $ \mathcal{A}^{\mu}_{\ \nu}$.

It is clear that, when evaluating $\Omega$ on two arbitrary horizontal lifts, $X$ and $Y$, we get
\begin{equation}
\Omega(X, Y) = g^{-1} \pi^{*} F(X, Y) g = g^{-1} F(\pi_{*} X, \pi_{*} Y) g = g^{-1} F(\tilde{X}, \tilde{Y}) g,
\label{curvatureOF}
\end{equation}
where $\tilde{X}$ and $\tilde{Y}$ are the projections of $X$ and $Y$ to $M$, respectively, and we use that $\text{i}_Y \text{i}_X F =F(X,Y)$. Finally, from equation \eqref{connMP}, we also have
\begin{equation}
0=\omega (\tilde{X}) = g^{-1} \mathcal{A} (X) g + g^{-1} \de_{P} g (\tilde{X}),
\label{diffeqlift}
\end{equation}
and
\begin{equation}
\de_{P} g (\tilde{X}) = \de g(X) =- \mathcal{A} (X) g .
\label{diffeqlift1}
\end{equation}
Equation \eqref{diffeqlift1} corresponds to a differential equation in terms of the horizontally lifted vector field, which plays an important role in the main body of the paper.

\bibliography{References}
\bibliographystyle{unsrt}

\end{document}